\RequirePackage{fix-cm}
\documentclass[twocolumn]{svjour3}          
\smartqed  
\usepackage{svg}

\usepackage{graphicx}
\usepackage{amsmath}
\usepackage{subcaption}
\usepackage{color}
\usepackage{amsmath,amssymb}
\usepackage[numbers,compress]{natbib}
\definecolor{someprettyred}{rgb}{0.85, 0.0, 0.1}

 \journalname{Granular Matter}

\begin{document}

\title{Granular binary mixtures improve energy dissipation efficiency of granular dampers}
\subtitle{}


\author{Nydia Roxana Varela-Rosales \and
        Angel Santarossa \and
        Michael Engel \and
        Thorsten P\"oschel 
}


\institute{Nydia Roxana Varela-Rosales \and Angel Santarossa \and  Michael Engel \and Thorsten P\"oschel \at
Institute for Multiscale Simulations\\
Friedrich-Alexander-Universit\"at Erlangen-N\"urnberg\\
Cauerstra\ss{}e 3, 91058 Erlangen\\
Germany\\
\email{thorsten.poeschel@fau.de}           
}

\date{Received: \today / Accepted: date}

\maketitle

\begin{abstract}
  Granular dampers are systems used to attenuate undesired vibrations produced by mechanical devices. They consist of cavities filled by granular particles. In this work, we consider a granular damper filled with a binary mixture of frictionless spherical particles of the same material but different size using numerical discrete element method simulations. We show that the damping efficiency is largely influenced by the composition of the binary mixture.


\end{abstract}

\section{Introduction}
\label{sec:intro}

Granular dampers are devices that exploit the dissipative nature of granular interactions to passively attenuate vibrations. In their most simple design, they consist of an enclosure partially filled by granular particles. When subjected to mechanical vibrations, the particles collide with each other and with the walls of the enclosure, thus, inelastic interactions (particle-particle and particle-wall) dissipate mechanical energy into heat. Granular dampers reveal distinctive characteristics beneficial for their application in technical devices: The design of granular dampers is simple and requires a minimum of maintenance, which makes them interesting for aerospace technologies \cite{Panossian1992} and applications in  weightlessness \cite{Sack2013,Sack:2015vh,sack2020granular}. They operate across a wide range of temperature \cite{Xia2019}, making them particularly useful in harsh environments.  Due to these features, granular dampers have numerous technical applications for instance in medicine \cite{Heckel2012}, construction \cite{Xu2004}, and in the vibration control during earthquakes \cite{Naeim,LU20122007,ZHOU2021113073}.

Granular dampers are subject to research for a long time, also using numerical discrete element method \linebreak (DEM) simulation techniques \cite{SPIE98,SPIE99,saluena1999}. In numerical studies and in experiments, it has been shown that the performance of granular dampers is sensitive to a variety of parameters such as the mass ratio between the granulate and the container \cite{wang2015}, number and material of the particles \cite{Hashemnia2021,els2011}, particle size \cite{Papalou_1998,Mao2001,Panossian1992} and shape \cite{pourtavakoli2016}, size and shape of the enclosure \cite{kollmer2013,meyer2021,ferreyra2021avoiding}, the free volume inside the container (gap clearance) \cite{wang2016,ZHANG2016}, and others \cite{ferreyra2021,lu2014,SANCHEZ20124389}. Yet, the effect of particle size dispersion on the efficiency of granular dampers was much less investigated. Only recently, the addition of micrometer-sized particles to a system of millimeter-sized monodisperse spheres was reported to have a significant effect on damping performance \cite{meyer2021}. An explanation of this effect, as well as the influence of the dispersion on damping efficiency in general remain open questions.

In this paper, we consider granular dampers using bidisperse granular mixtures by means of DEM simulation, which is reliable numerical tool to study granular systems \cite{Algo,Matuttis:2014}.

\section{System setup and numerical model}
\label{sec1_2_methods}

Our damper of cubical shape ($8\times8\times8\text{ cm}^3$) is filled partially by $N$ frictionless spheres. For both container and particles, we assume the same material characteristics corresponding to steel (density $\rho = 7,850\, \text{kg/m}^3$, Young modulus $E = 2\times10^{9}\,\text{Pa}$,  Poisson ratio $\nu=0.3$).

\begin{figure}
  \centering
  \includegraphics[width=1.0\linewidth]{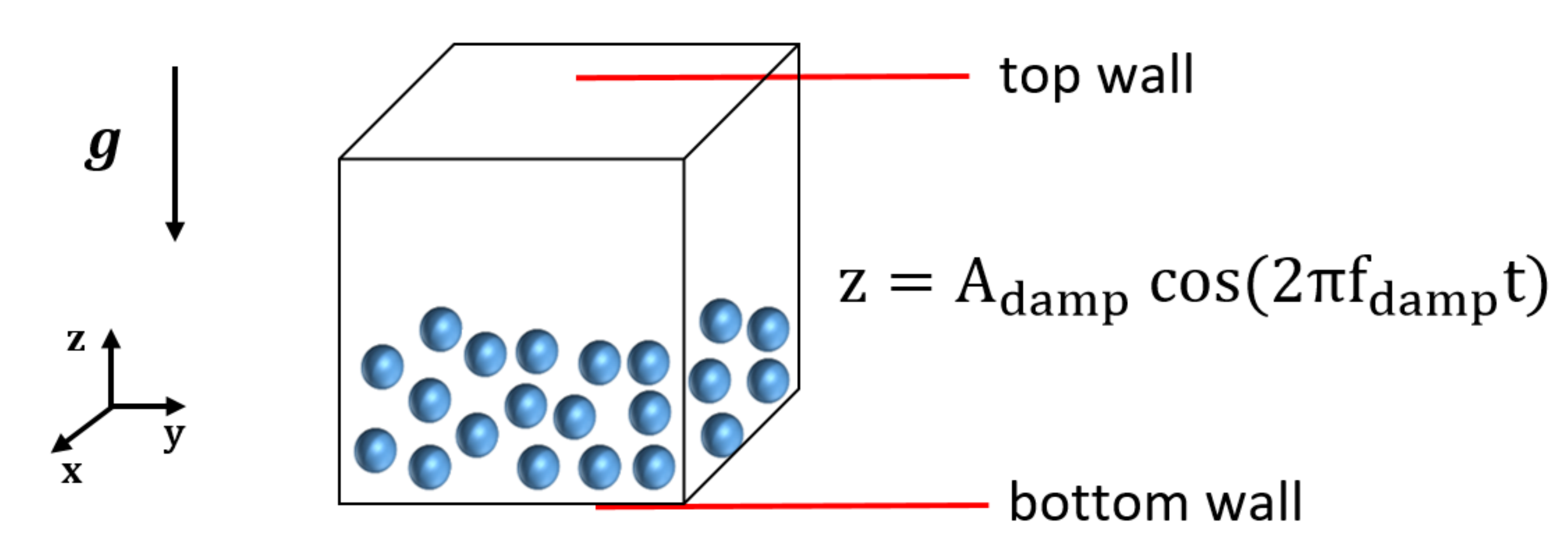}
  \caption{Sketch of the granular damper. The walls of
the enclosure in $z$ direction are referred to as top and bottom walls, respectively.}
  \label{fig:sketch}
\end{figure}

The damper is driven by oscillations, \linebreak $z(t)=A_\text{damp} \cos(2 \pi f t)$ with $f_\text{damp} = 70\,\text{Hz}$ where gravity, $g = 9.8\,\text{m/s}^2$, acts in negative $z$ direction. Figure \ref{fig:sketch} shows a sketch of the setup.

The interaction force between viscoelastic frictionless spheres is given by the elastic Hertz law \cite{Hertz:1881} and an appropriate dissipative force. The Hertz law results from the assumption that the elastic stress is a linear function of the strain, therefore, for the dissipative force it is reasonable to assume that the dissipative stress is a linear function of the strain rate \cite{BSHP}. Combining these forces, we obtain the absolute value of the force between particles $i$ and $j$ of radii $R_i$ and $R_j$ at positions $\vec{r}_i$ and $\vec{r}_j$ at velocities $\vec{v}_i$ and $\vec{v}_j$.  When in contact, that is, if $\xi_{ij}\equiv R_i+R_j - \left|\vec{r}_i-\vec{r}_j\right|\ge 0$, the force is
\begin{equation}
  \label{eq:FN}
  F_{ij}= \frac{2E\sqrt{R^\text{eff}_{ij}}}{3(1-\nu^2)} \left(\xi^{3/2} + A\sqrt{\xi}\,\dot{\xi}\right)
\end{equation}
\begin{equation}
  \label{eq:FNvec}
  \vec{F}_{ij}=\max\left(0,\, F_{ij}\right)\,\frac{\vec{r}_i-\vec{r}_j}{\left|\vec{r}_i-\vec{r}_j\right|}\,,
\end{equation}
with $1/R_{ij}^\text{eff}\equiv 1/R_i + 1/R_j$. The dissipative constant, $A$, is a function of the viscosities of the particles' materials \cite{BSHP}, which are difficult to determine. A much easier way to determine $A$ is through the coefficient of restitution at a certain impact rate \cite{MuellerPoeschel:2011}. For our simulations, to determine $A$, we assume the coefficient of restitution $\varepsilon=0.75$ for steel particles colliding at impact rate 1 m/s. For bidisperse mixtures, we obtain $A$ for the collision of particles of each species and calculate its arithmetic mean. In some cases, it can happen that the dissipative force overcompensates the restoring elastic force, resulting in a total attractive force (see \cite{Algo} for a detailed explanation). The maximum rule in Eq. \ref{eq:FNvec} ensures that the interaction force is always repulsive. In our force model, we neglect frictional forces between contacting particles, justified by previous evidence  \cite{Bai2009,Bannerman2011,pourtavakoli2016}, which shows that the main dissipation mechanism in granular dampers is due to normal interaction forces. For our DEM simulations, we use Yade \cite{vsmilauer2021yade-3}.

\section{Energy dissipation per cycle of oscillation}

We quantify energy dissipation efficiency by

\begin{equation}
  \label{Ediss}
  {E_\text{d}(t) \equiv  \frac{(E_\text{pp} + E_\text{pw})}{E_\text{max}}      
  }\,,
\end{equation}
where $E_\text{pp}$ is given by
\begin{equation}
  \label{Epp}
  {E_\text{pp}(t) = \frac{1}{2T} 
  \int_{t}^{t+T} \sum\limits_{i=1}^N \sum\limits_{i \neq j}^N
  \vec{F}_{ij}\cdot \left(\vec{v}_j-\vec{v}_i\right) dt }\,,
\end{equation}
and $E_\text{pw}$ by
\begin{equation}
  \label{Epw}
   {E_\text{pw}(t) = \frac{1}{T} 
  \int_{t}^{t+T} \sum\limits_{i=1}^N \sum\limits_{w=1}^M \vec{F}_i^w\cdot \vec{v}_i \;  dt }\,,
\end{equation}
with $T=1/f$. Equation \eqref{Ediss} relates the energy dissipated by particle-particle ($E_\text{pp}$) and particle-wall ($E_\text{pw}$) contacts during one period $T$ of oscillation to the maximum energy, $E_\text{max}$, that can be dissipated in one oscillation cycle.
The quantity $E_\text{max}$ was defined in \cite{Sack2013}. It is obtained by assuming that all particles collide twice per cycle with the bottom and top walls of the container at maximal relative velocity and dissipate the energy of the relative motion in each of these collisions.

Note that $E_\text{d}$ is similar but not identical to the quantity $b$ defined in Eq. (4) of \cite{saluena1999}, which relates the dissipated energy to the total kinetic energy. 

The dissipated energy depends on the dynamics of the granulate and is, thus, a fluctuating quantity. Therefore, we compute an average, $\left<E_\text{d}\right>$, from a linear fit to $E_\text{d}(t)$. 

\section{Modes of operation of granular dampers}
\label{sec:modes-oper-class}

When operating in weightlessness, depending on the amplitude, $A_\text{damp}$, of the oscillation, granular dampers reveal different modes  \cite{Sack2013}. For small $A_\text{damp}$, the granular material exhibits gas-like behavior. In this state, only a small fraction of the kinetic energy is dissipated. More relevant for granular damping is the \emph{collect-and-collide} regime \cite{Bannerman2011} observed for large $A_\text{damp}$, where the center of mass of the granulate moves synchronously with the container. In this state, during the inward stroke, the granulate accumulates at the wall of the container forming a packed layer. When the acceleration of the box decreases, this layer leaves the wall collectively and eventually collides with the opposing wall where a large fraction of the kinetic energy is dissipated.
The modes of operation are separated by a sharp transition at a critical value of $A_\text{damp}$; here the dissipated energy per oscillation achieves its maximum value \cite{Sack2013,sack2020granular}. Motivated by this argument, in the results section, we consider the interval $A_\text{damp}\in [2,4]\,\text{cm}$ containing the critical value.

A more detailed classification of modes of behavior of vibrated boxes filled by granular material was presented in \cite{Opsomer:2011}, however, not in the context of granular damping.

\section{Results}\label{sec2_results}

\subsection{Classes of equal binary mixtures}
\label{sec:classes-mixtures}

We consider three classes of binary mixtures in comparison with a monodisperse reference system:
(a) the total mass of the granulate ($M_\text{tot}$) in the bidisperse system equals the mass in the monodisperse reference system, but not the number of particles;
(b) the total number of particles ($N_\text{tot}$) in the bidisperse system equals the number of particles in the reference system, but not the total mass of the granulate;
(c) both the number of particles and the mass of the granulate in the bidisperse system equal the corresponding quantities in the monodisperse reference system, thus, isolating the effect of size dispersion from particle number and mass. We summarize the properties of the classes (a)-(c) in Tab. \ref{tab:classes}.
\begin{table}[htbp]
\centering
\begin{tabular}{l|l|l|p{1.3cm}}
& large & small & quantities \\
& particles & particles & conserved\\
  \hline
  reference & \multicolumn{3}{l}{$N_\text{ref}=300$, $R_\text{ref} = 0.25\,\text{cm}$, $M_\text{ref}=0.154\text{kg}$}\\
  system \\
  \hline
  class (a) & $N_\text{tot} / 2$ & $N_\text{tot} / 2$ & \mbox{$M_\text{tot}=M_\text{ref}$} \\
  $\sigma\in[0.1,0.9]$ &$R=R_\text{ref}$ & $R=\sigma R_\text{ref}$ \\
  \hline
  class (b)  &  $N_\text{ref}/2$ & $N_\text{ref}/2$ & \mbox{$N_\text{tot}=N_\text{ref}$} \\
 $\sigma\in[0.1,0.9]$ &$R=R_\text{ref}$ & $R=\sigma R_\text{ref}$\\
  \hline
  class (c)  & $N_\text{ref}/2$ & $N_\text{ref}/2$ & \mbox{$N_\text{tot}=N_\text{ref}$}  \\
  $\sigma\in[0.1,0.9]$ &                         $R=\hat{\sigma} R_\text{ref}$ &  
  $R=\sigma R_\text{ref}$  &
  \mbox{$M_\text{tot}=M_\text{ref}$} \\
  $\hat{\sigma}^3=2-\sigma^3$ &  & \\ 
\end{tabular}

  \caption{Classes of binary mixtures studied in this work.}
  \label{tab:classes}
\end{table}

\subsection{Class (a): Equal total mass}
\label{SC_1}

\begin{figure}
	\centering
	\includegraphics[width=1.0\linewidth]{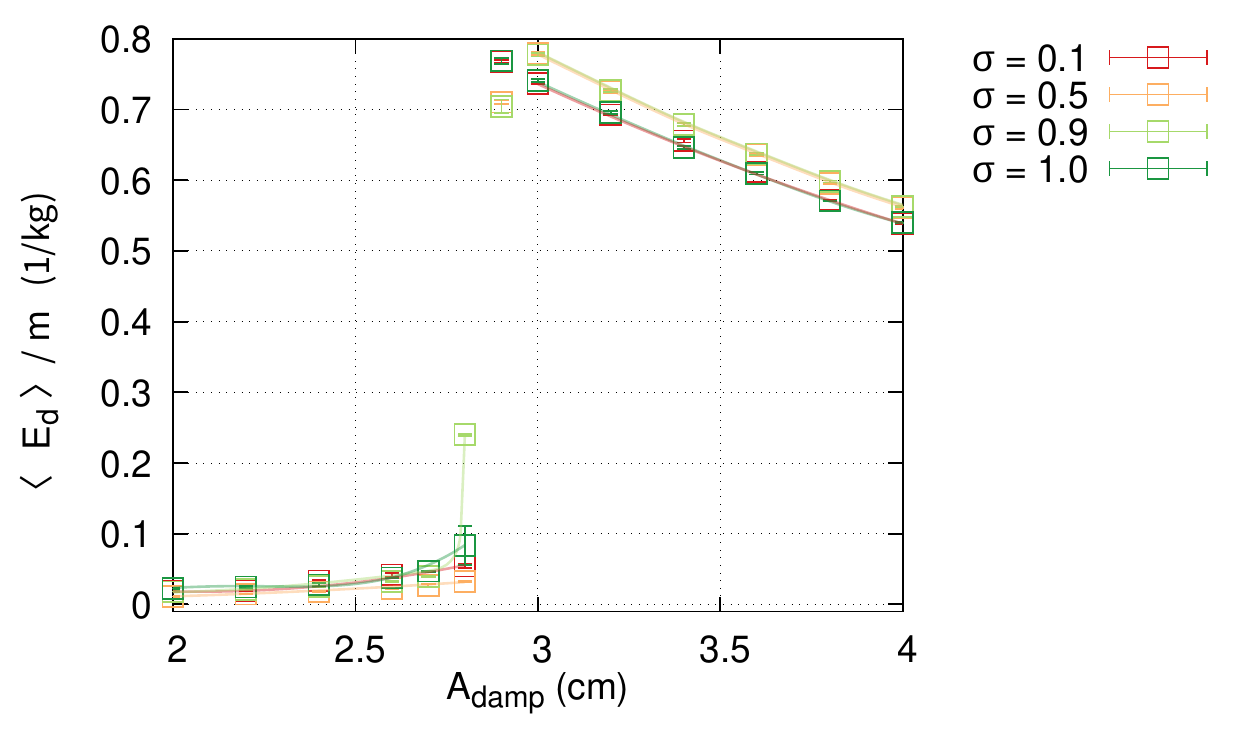}
	\caption{Class (a): Energy dissipation per cycle of oscillation, $\left<E_\text{d}\right>$, normalized by the mass, $m$, of the granulate, as a function of the driving amplitude. For each value of $\sigma$, the data was averaged over 50 periods. Error bars show the standard deviation.}
	\label{fig:dissE_CM}
\end{figure}

Figure \ref{fig:dissE_CM}
 shows the  average energy dissipation efficiency per cycle, $\left<E_\text{d}\right>$, normalized by the total mass, $m$, of the
granulate, as a function of the vibration amplitude for different size ratios, $\sigma$. For all values of $\sigma$, the system reveals a similar behavior. An optimal value of the driving amplitude, $A_\text{opt}$, divides the dynamics into two regimes: For $A_\text{damp}\lesssim 2.8\,\text{cm}$,  the granulate is in a gaseous state where only a small fraction of the particles collides with the walls during an oscillation period. In contrast, for $A_\text{damp}\gtrsim 2.8\,\text{cm}$, the center of mass of the granulate moves synchronously with the external driving, in the collect-and-collide regime. Both modes of dynamic behavior, gaseous and collect-and-collide, and their dissipative properties are discussed in detail in \cite{Sack2013,Sack:2015vh,sack2020granular,Bannerman2011}.

\begin{figure}
	\centering
	\includegraphics[width=.48\textwidth]{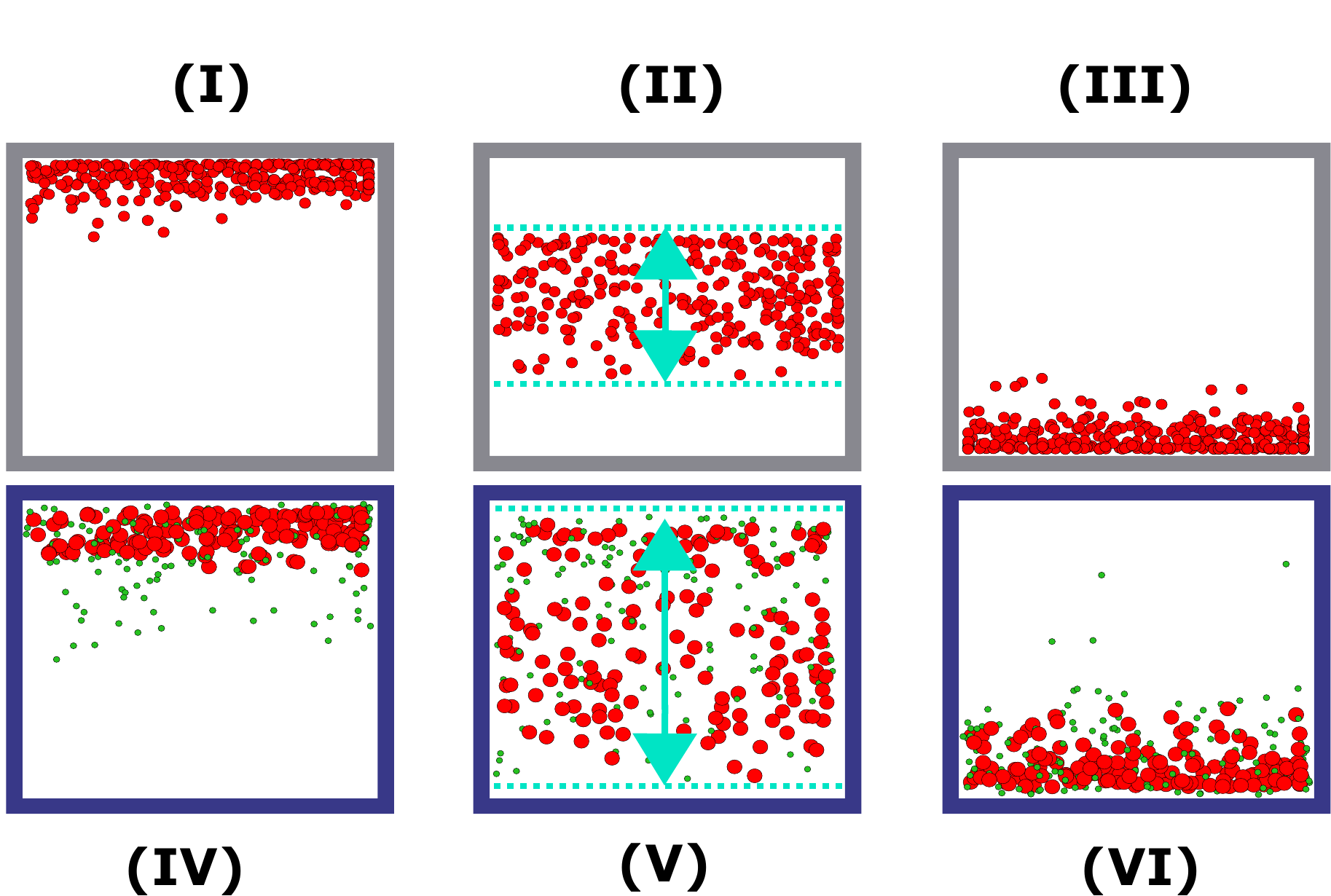}
	\caption{Class (a): Snapshots of the monodisperse reference system (grey boxes) at the times of minimal (I),(III) and maximal (II) velocity and the same for a bidisperse granulate blue boxes with $\sigma = 0.1$ (IV)-(VI). Dotted lines in (II), (V) indicate the region populated by particles. Red circles show particles with radius $R = R_\text{ref}$,  green circles show small particles, $R=\sigma\,R_\text{ref}$.}
	\label{fig:sim-snapshots}
\end{figure}

Figure \ref{fig:sim-snapshots}
shows snapshots of the simulation at the times of minimal and maximal velocity for the reference system and for a bidisperse system with $\sigma=0.1$ in the collect-and-collide regime. We notice that for bidisperse systems the particles spread over a larger volume in the container. Moreover, we observe size segregation. 

For $A_\text{damp} > A_\text{opt}$ (Fig. \ref{fig:dissE_CM}), the bidisperse system with $\sigma = 0.5$ achieves higher values of $\left<E_\text{d}\right>/m$ than the reference system, i.e., it dissipates energy more efficiently than the monodisperse reference system, in the collect-and-collide mode.

\begin{figure}
\centering
    \includegraphics[width=.5\textwidth]{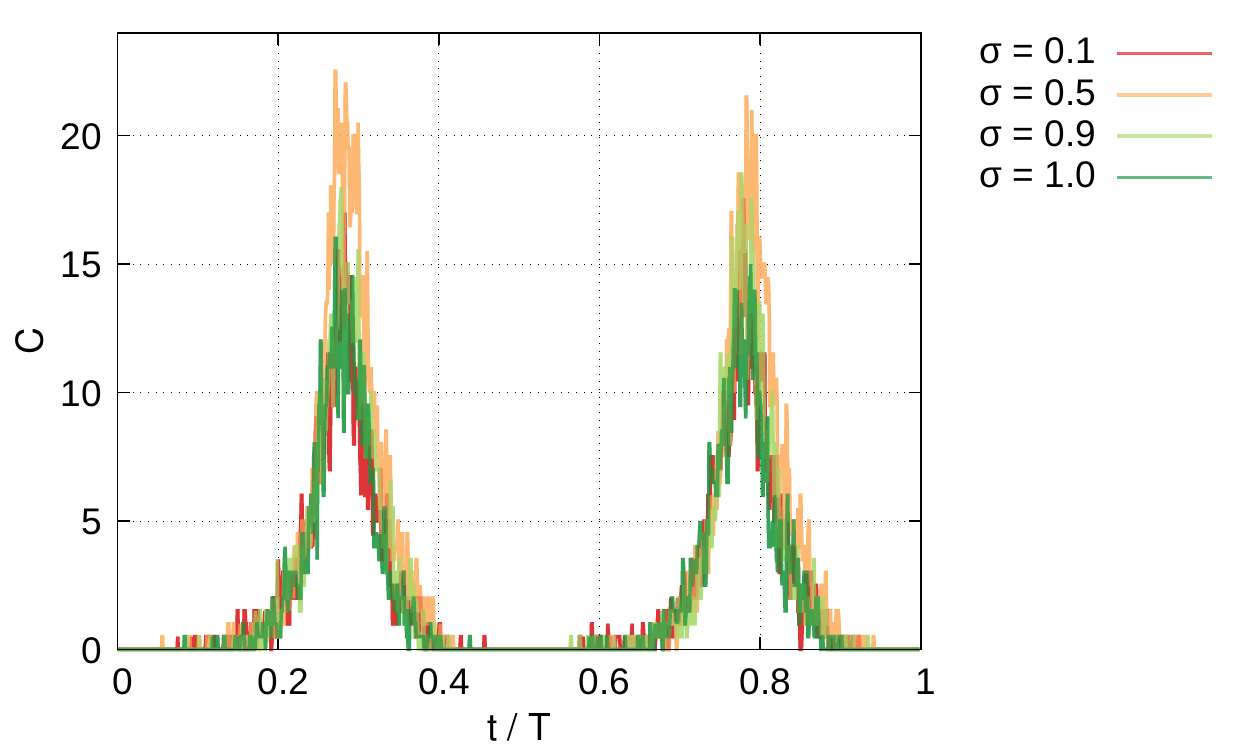}
    \caption{Class (a): Number of particles $C$ in contact with the bottom or top wall as a function of time, for $A_\text{damp}= 3\,\text{cm}$. For each value of $\sigma$, the data was averaged over 50 periods.}
\label{fig:CM-collions}
\end{figure}

To understand this behaviour, we consider the particles' collisions with the top and bottom walls, which dominate the dissipation \cite{Bai2009}. In each time step of the integration, we record the number of particles contacting one of these walls (Fig. \ref{fig:CM-collions}).
We note that for bidisperse systems, the collision frequency is higher than for the reference system ($\sigma=1$). Detailed analysis shows that this difference can be attributed to collisions of small particles with the walls.

\begin{figure}
\centering
    \includegraphics[width=.5\textwidth]{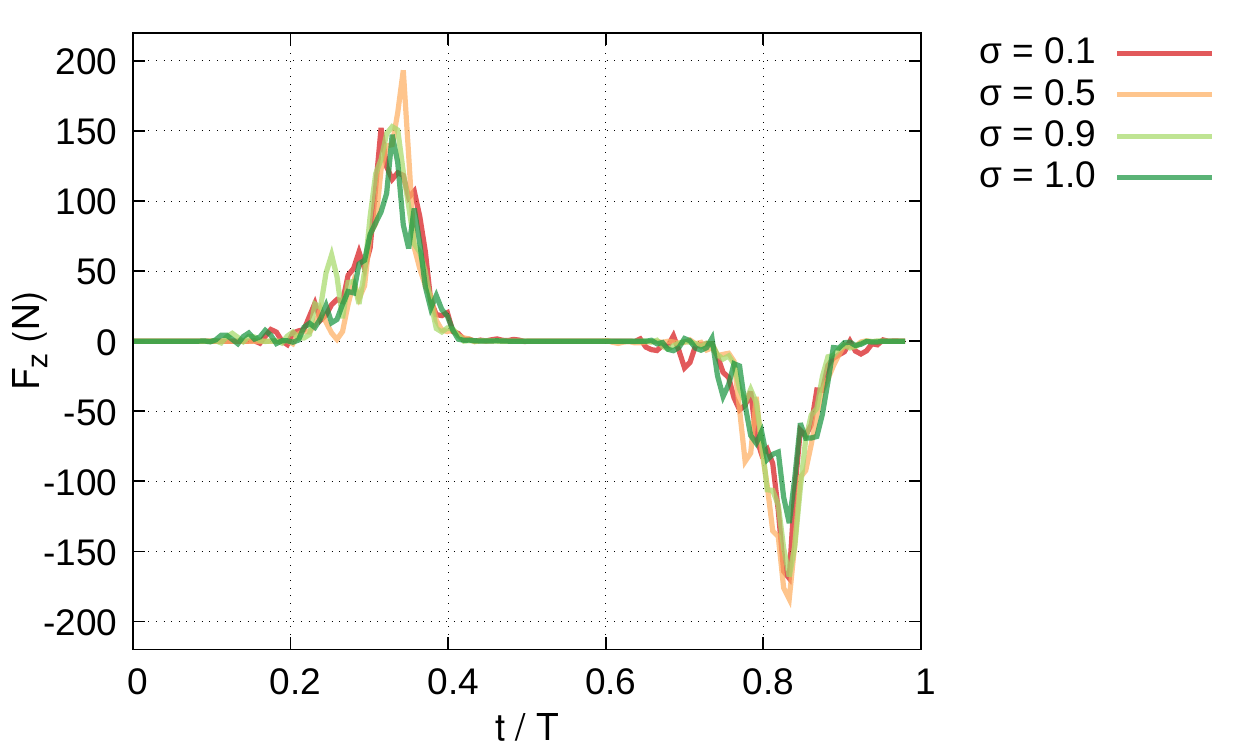}
    \caption{Class (a): Force in $z$ direction exerted on the bottom and top wall as a function of time, for $A_\text{damp}= 3\,\text{cm}$. For each value of $\sigma$, the data was averaged over 50 periods.}
\label{fig:CM-force}
\end{figure}

The number of particle contacts alone is, however, not sufficient to explain the difference in energy dissipation. Therefore, we record the force in $z$ direction ($F_z$) exerted by the granulate on the top and bottom walls (Fig. \ref{fig:CM-force}). The bidisperse damper with $\sigma=0.5$ show a higher value of $F_z$, compared to the reference system. Higher absolute values of $F_z$ lead to higher momentum transfer and, in turn,  to larger loss of kinetic energy in dissipative collisions, which explains the difference in the energy dissipation efficiency shown in Fig. \ref{fig:dissE_CM}.

\subsection{Class (b): Equal number of particles}
\label{CN}

\begin{figure}
  \centering
  \includegraphics[width=.5\textwidth]{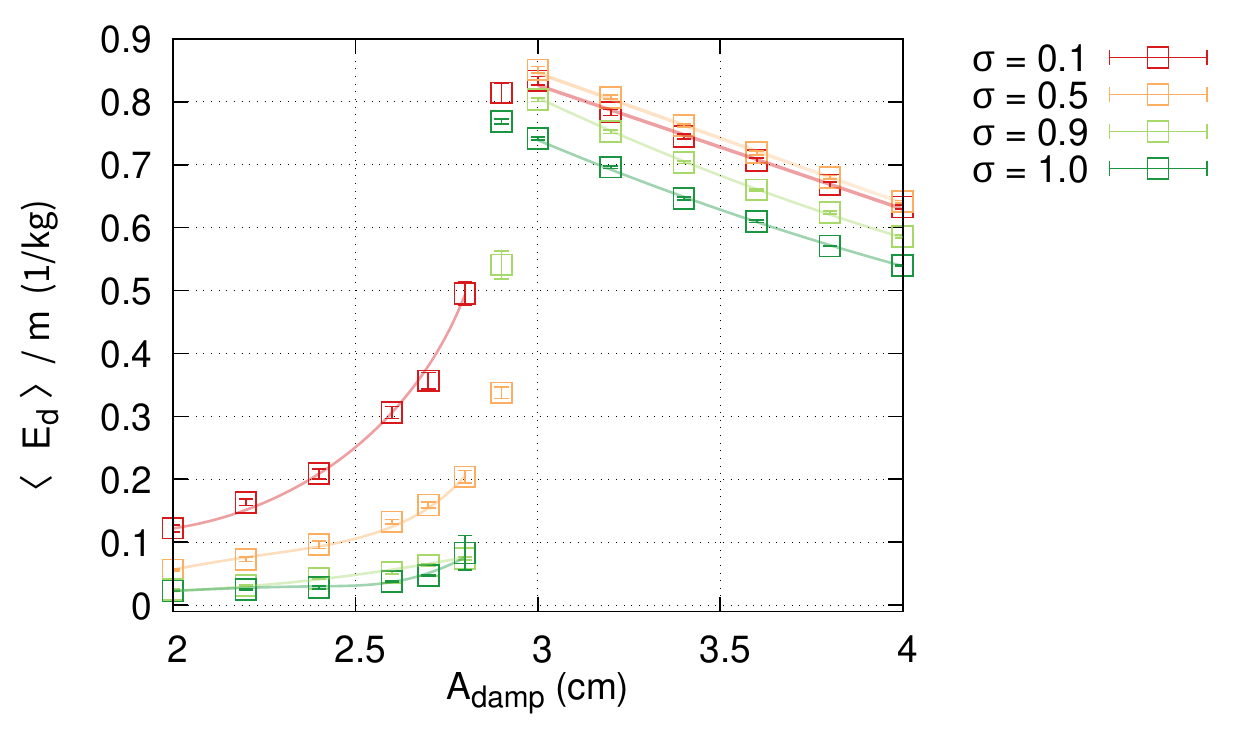}
  \caption{Class (b): Energy dissipation per cycle of oscillation, $\left<E_\text{d}\right>$, normalized by the mass, $m$, of the granulate, as a function of the driving amplitude. For each value of $\sigma$, the data was averaged over 50 periods. Error bars show the standard deviation.}
  \label{fig:dissE_CN}
\end{figure}

Figure \ref{fig:dissE_CN}
shows the average energy dissipation efficiency per cycle, $\left<E_\text{d}\right>$, normalized by the total mass, $m$, of the
granulate, as a function of the vibration amplitude for different size ratios, $\sigma$, corresponding to Fig. \ref{fig:dissE_CM} for class (a) binary mixtures. For all values of $\sigma$, the systems exhibit a similar behavior. Again, we see two regimes for small and large amplitude (gaseous and collect-and-collide) and again dissipation efficiency for bidisperse mixtures exceeds that of the reference system ($\sigma=1$).

\begin{figure}
  \centering
  \includegraphics[width=.5\textwidth]{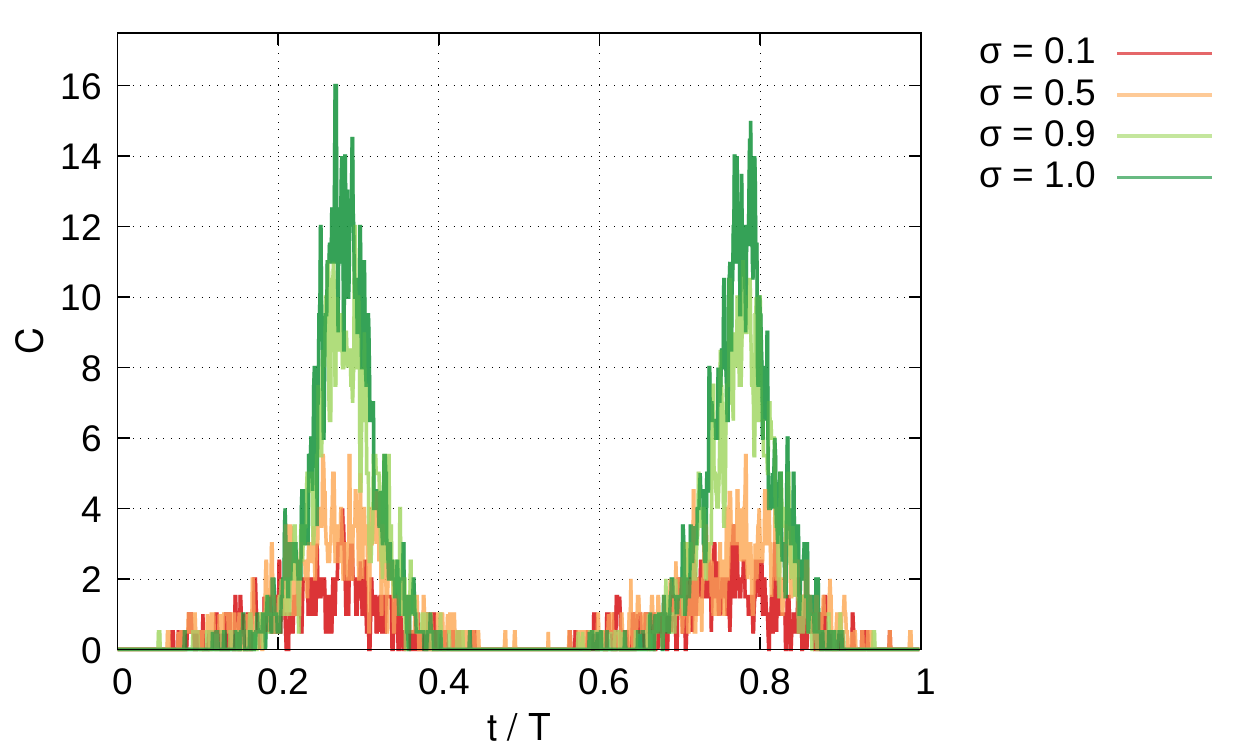}
  \caption{Class (b): Number of particles $C$ in contact with the bottom or top wall as a function of time, for $A_\text{damp}= 3\,\text{cm}$. For each value of $\sigma$, the data was averaged over 50 periods.}
  \label{fig:CN-collions}
\end{figure}

From Fig. \ref{fig:CN-collions}
 we observe that the number of contacts with the top and bottom walls of the reference system ($\sigma=1$) exceeds that of mixtures for all values $\sigma<1$. Consequently, the difference of the dissipation shown in Fig. \ref{fig:dissE_CN} cannot be attributed to the number of contacts.

\begin{figure}
  \includegraphics[width=.5\textwidth]{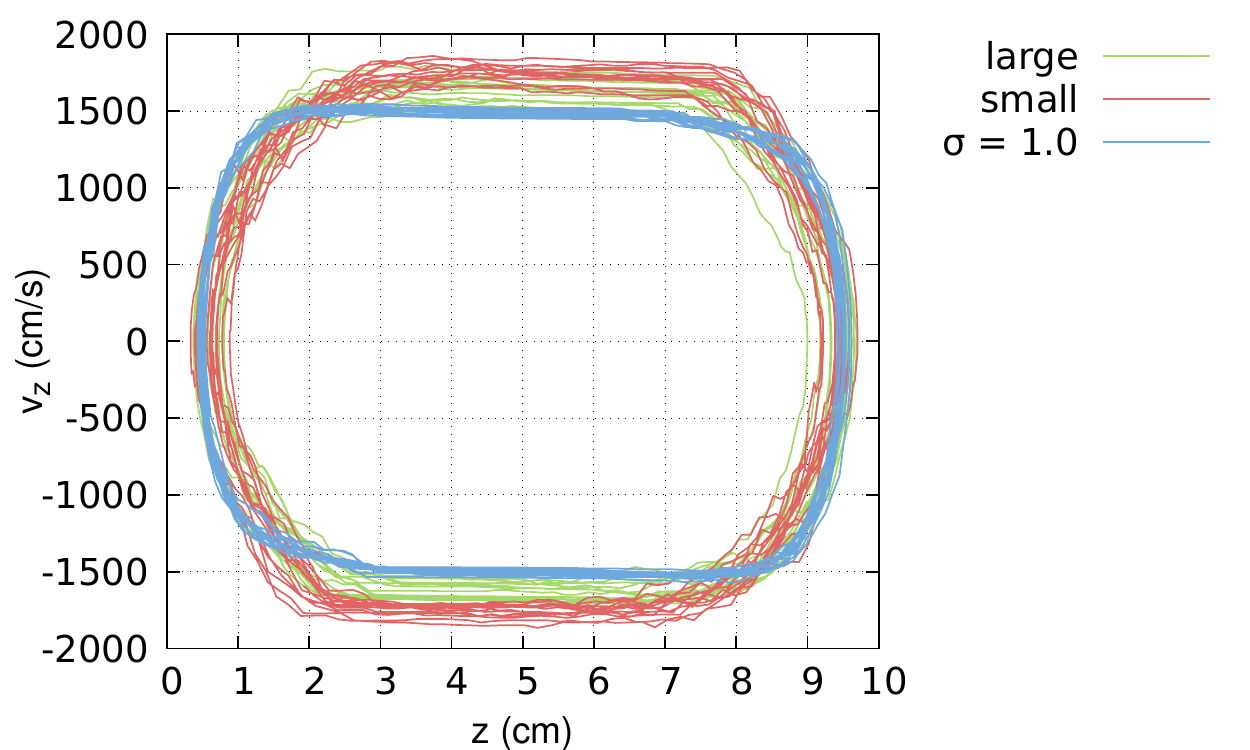}
  \caption{Class (b): Phase plot, $v_\text{z} (z)$, of the $z$-component of the center-of-mass position for $\sigma = 1$ (reference) and $\sigma = 0.5$. Amplitude is $A_\text{damp} = 3\,\text{cm}$. For the bidisperse system, the plot is performed separately for small and large particles.}
  \label{fig:CN-md}
\end{figure}

From the phase plot of the center-of-mass position in the direction of the oscillation, $z$, (Fig. \ref{fig:CN-md}),
we see that both small and large particles of the bidisperse system (with $\sigma=0.5$) assume larger velocities than the particles of the reference system ($\sigma=1$). Higher absolute velocities give rise to higher impact velocities and, thus, larger loss of kinetic energy in dissipative collisions, which explains the difference in the damping efficiency shown in Fig. \ref{fig:dissE_CN}.

\subsection{Class (c): Equal total mass and number of particles}

\begin{figure}
  \centering
  \includegraphics[width=.5\textwidth]{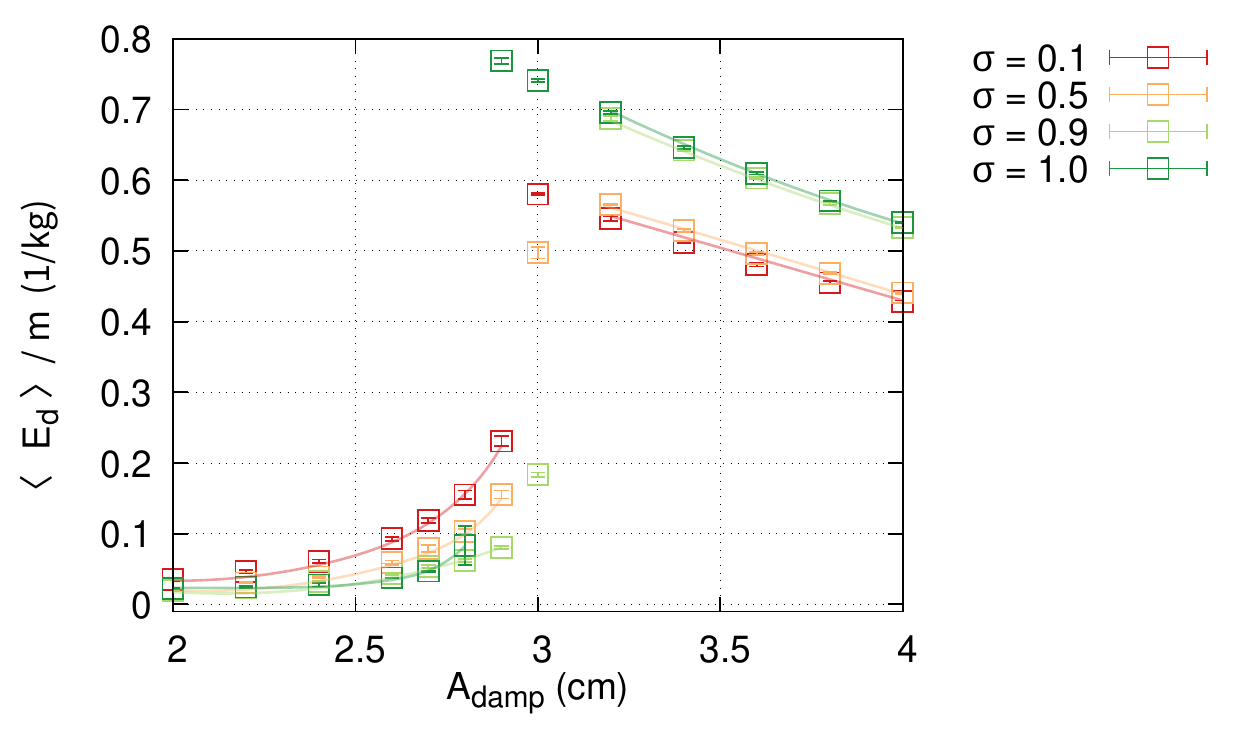}
  \caption{Class (c): Energy dissipation per cycle of oscillation, $\left<E_\text{d}\right>$, normalized by the mass, $m$, of the granulate, as a function of the driving amplitude. For each value of $\sigma$, the data was averaged over 50 periods. Error bars show the standard deviation. In contrast to classes (a) and (b), here we find more efficient dissipation for the reference system ($\sigma=1$) than for bidisperse systems.}
	\label{fig:dissE_CMCN}
\end{figure}

\begin{figure}
  \includegraphics[width=.5\textwidth]{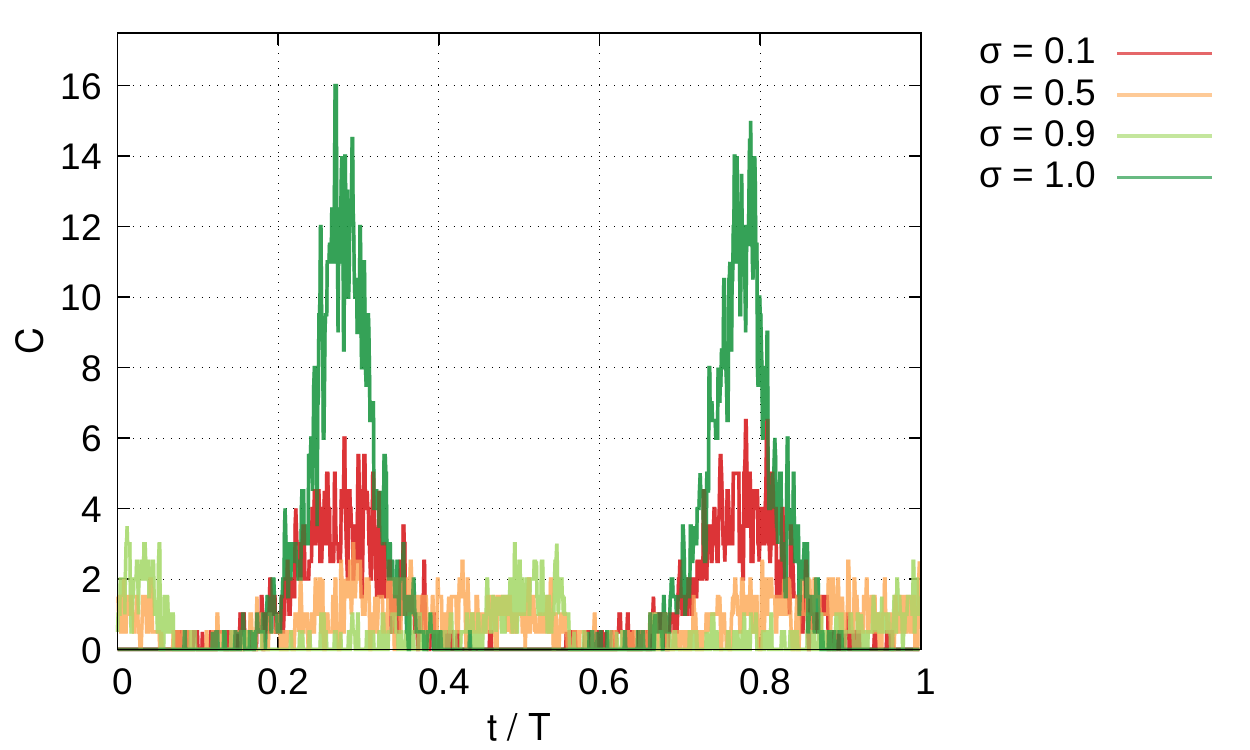}
  \caption{Class (c): Number of particles $C$ in contact with the bottom or top wall as a function of time, for $A_\text{damp}= 3\,\text{cm}$. For each value of $\sigma$, the data was averaged over 50 periods.}
  \label{fig:CMCN-collions}
\end{figure}

\begin{figure}
\centering
    \includegraphics[width=.5\textwidth]{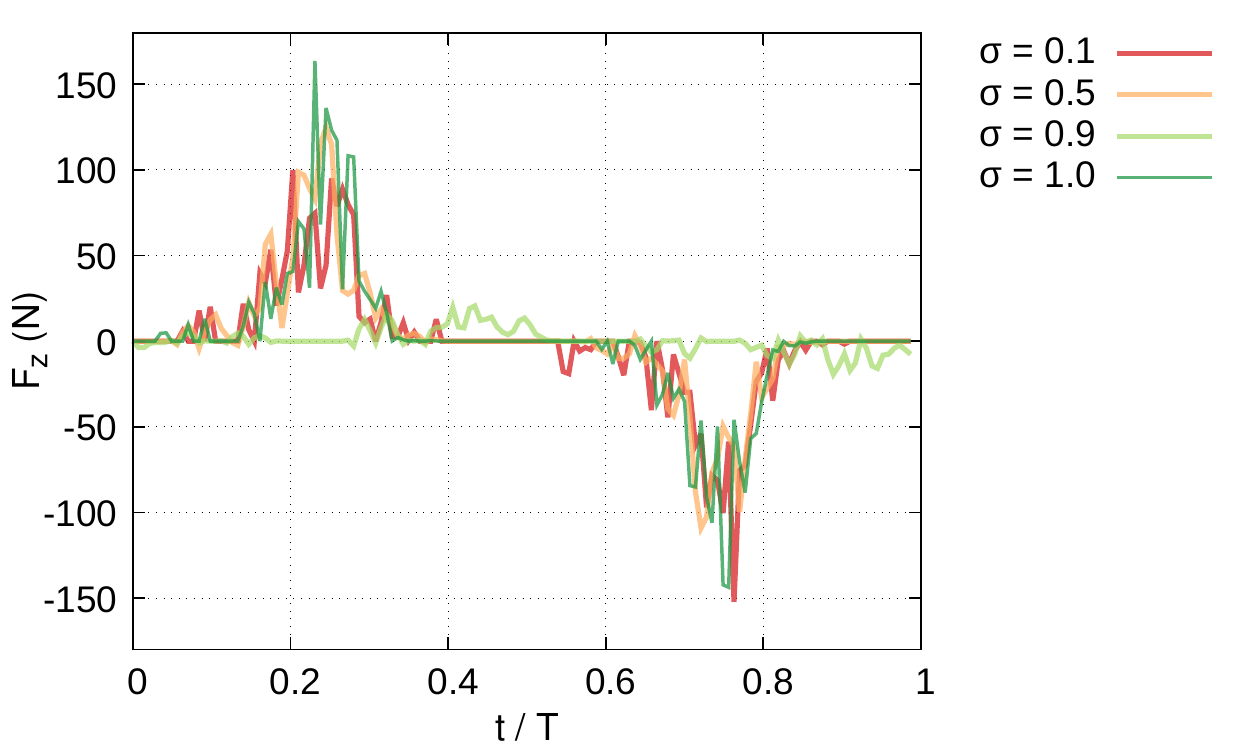}
    \caption{Class (c): Force in $z$ direction exerted on the bottom and top wall as a function of time, for $A_\text{damp}= 3\,\text{cm}$. For each value of $\sigma$, the data was averaged over 50 periods.}
\label{fig:CMCN-force}
\end{figure}

Similar as for the previous classes, also for class (c) $\left<E_\text{d}\right>/m$, as a function of $A_\text{damp}$ (Fig. \ref{fig:dissE_CMCN}),
shows two well separated regimes for large and small $A_\text{damp}$. In contrast to classes (a) and (b), here the transition value of $A_\text{damp}$ depends on $\sigma$. Also in contrast to classes (a) and (b), here the values of $\left<E_\text{d}\right>/m$ of monodisperse reference system for $A_\text{damp} > 2.8\,\text{cm}$ exceed the values of dampers with $\sigma<1$. Here, the number of contacts (Fig. \ref{fig:CMCN-collions})
of the reference system exceeds the value for the granular mixtures. Furthermore, the force in $z$ direction, $F_z$, (Fig. \ref{fig:CMCN-force}), exerted by the particles of the reference system is considerably larger than for bidisperse dampers.

\begin{figure}
  \includegraphics[width=.5\textwidth]{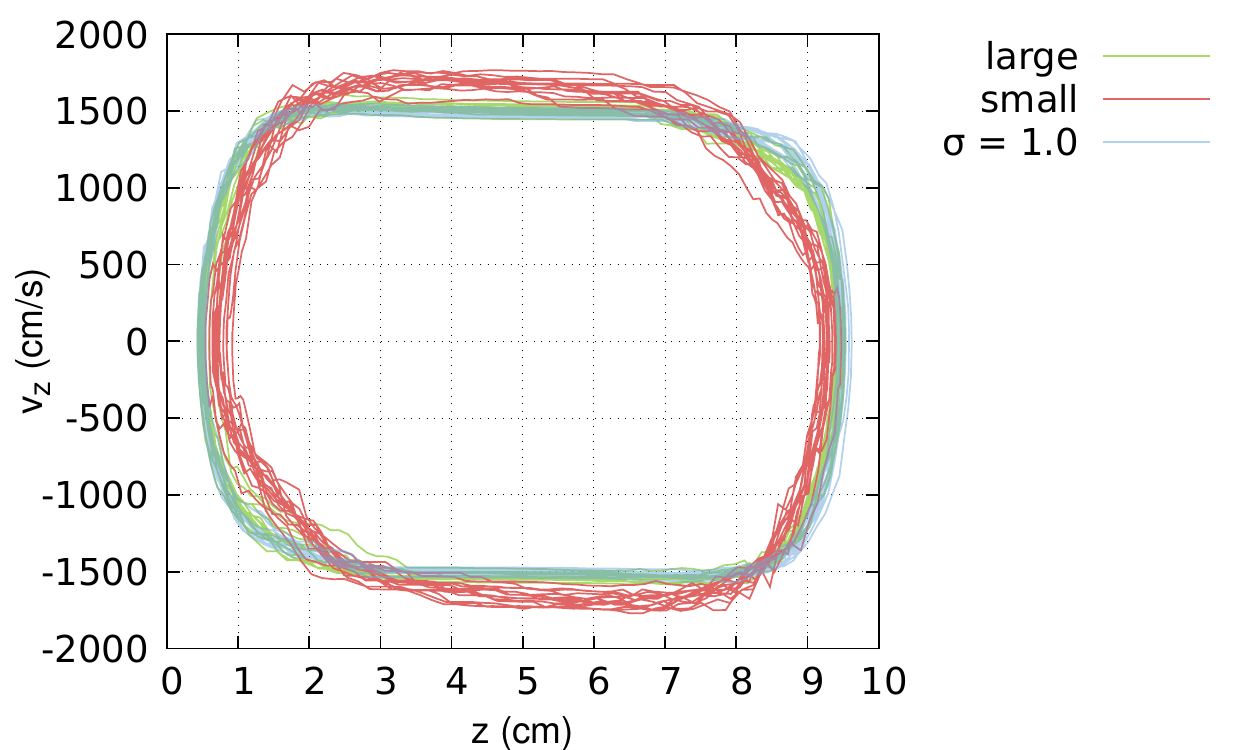}
  \caption{Class (c): Phase plot, $v_\text{z} (z)$, of the $z$-component of the center-of-mass position for $\sigma = 1$ (reference) and $\sigma = 0.5$. Amplitude is $A_\text{damp} = 3\,\text{cm}$. For the bidisperse system, the plot is performed separately for small and large particles.}
  \label{fig:CMCN-md}
\end{figure}

From the phase plot for $\sigma=0.5$, Fig. \ref{fig:CMCN-md},
we see that the large particles follow approximately the same path as the particles of the reference system while the small particles achieve larger velocities. Large and small particles in the bidisperse system do not move collectively. In contrast to dampers of class (b), there is no energy dissipation enhancement due to a higher center of mass velocity of both large and small particles.

\section{Conclusion}\label{sec13}

The efficiency of granular dampers operating with a bidisperse granular mixture depends sensitively on the composition of the mixture. Thus, the choice of the composition can enhance energy dissipation in granular dampers significantly as compared to granular dampers operating with a monodisperse granulate of the same material. For the comparison with a monodisperse reference system, we considered  three classes of mixtures: (a) both monodisperse and bidisperse systems are of the same total mass of the granulate (but different number of particles); (b) both systems have the same total number of particles (but different mass); (c) the sizes of the particles in the mixture are scaled such that both systems contain the same number of particles of  the same total mass.

For dampers of class (a), the efficiency increases due to a larger frequency of collisions and higher exerted force during impact with the top and bottom walls in comparison with the reference system. 

For dampers of class (b), the efficiency increases due to the larger relative velocity between the granular particles and the wall of the container, shown by the phase plot of the center of mass velocity of the binary system in comparison with the monodisperse reference system. 

For dampers of class (c), the energy dissipation efficiency is smaller than for the monodisperse reference system due to a smaller frequency of collisions and lower exerted force during impact with the top and bottom walls. Furthermore, there is no energy dissipation improvement due to a higher center of mass velocity of both large and small particles.

In conclusion, we find that the dissipative properties of mixtures of differently sized particles differ from the dissipative properties of monodisperse systems. Therefore, the particle size distribution of the granulate is an important feature for the design of granular dampers.

\begin{acknowledgements}
The work was supported by the Interdisciplinary Center for Nanostructured Films (IZNF), the Competence Unit for Scientific Computing (CSC), and the Interdisciplinary Center for Functional Particle Systems (FPS) at Friedrich-Alexander University Erlangen-N\"urnberg. N. R. V.-R. and M. E. acknowledge funding by the Deutsche \linebreak 
Forschungsgemeinschaft (DFG) through project number 
\linebreak 406658237. A. S. and T. P. acknowledge funding by DFG under project number 411517575 and through the Research Training Group GRK 2423 ``Fracture across Scales –\textsc{Frascal}''.
\end{acknowledgements}

\section*{Compliance with ethical standards}

The authors declare that they have no conflict of interest.

\bibliography{granularDamper.bib}   

\begin{thebibliography}{36}
\providecommand{\natexlab}[1]{#1}
\providecommand{\url}[1]{\texttt{#1}}
\expandafter\ifx\csname urlstyle\endcsname\relax
  \providecommand{\doi}[1]{doi: #1}\else
  \providecommand{\doi}{doi: \begingroup \urlstyle{rm}\Url}\fi

\bibitem[Panossian(1992)]{Panossian1992}
H.~V. Panossian.
\newblock Structural damping enhancement via non-obstructive particle damping
  technique.
\newblock \emph{J. Vib. Acoust.}, 114:\penalty0 101--105, 1992.
\newblock \doi{10.1115/1.2930221}.

\bibitem[Sack et~al.(2013)Sack, Heckel, Kollmer, Zimber, and
  P\"oschel]{Sack2013}
A.~Sack, M.~Heckel, J.~E. Kollmer, F.~Zimber, and T.~P\"oschel.
\newblock Energy dissipation in driven granular matter in the absence of
  gravity.
\newblock \emph{Phys. Rev. Lett.}, 111:\penalty0 018001, 2013.
\newblock \doi{10.1103/PhysRevLett.111.018001}.

\bibitem[Sack et~al.(2015)Sack, Heckel, Kollmer, and P{\"o}schel]{Sack:2015vh}
A.~Sack, M.~Heckel, J.~E. Kollmer, and T.~P{\"o}schel.
\newblock Probing the validity of an effective-one-particle description of
  granular dampers in microgravity.
\newblock \emph{Granul. Matter}, 17:\penalty0 73--82, 2015.
\newblock \doi{10.1007/s10035-014-0539-8}.

\bibitem[Sack et~al.(2020)Sack, Windows-Yule, Heckel, Werner, and
  P{\"o}schel]{sack2020granular}
A.~Sack, K.~Windows-Yule, M.~Heckel, D.~Werner, and T.~P{\"o}schel.
\newblock Granular dampers in microgravity: sharp transition between modes of
  operation.
\newblock \emph{Granul. Matter}, 22:\penalty0 54, 2020.
\newblock \doi{10.1007/s10035-020-01017-x}.

\bibitem[Xia et~al.(2009)Xia, Liu, Shan, and Li]{Xia2019}
Z.~Xia, X.~Liu, Y.~Shan, and X.~Li.
\newblock Coupling simulation algorithm of discrete element method and finite
  element method for particle damper.
\newblock \emph{J. Low Freq. Noise Vib. Act. Control}, 28:\penalty0 197--204,
  2009.
\newblock \doi{10.1260/026309209790252545}.

\bibitem[Heckel et~al.(2012)Heckel, Sack, Kollmer, and P{\"o}schel]{Heckel2012}
M.~Heckel, A.~Sack, J.~E. Kollmer, and T.~P{\"o}schel.
\newblock Granular dampers for the reduction of vibrations of an oscillatory
  saw.
\newblock \emph{Physica}, A391:\penalty0 4442--4447, 2012.
\newblock \doi{10.1016/j.physa.2012.04.007}.

\bibitem[Xu et~al.(2004)Xu, Wang, and Chen]{Xu2004}
Z.~Xu, M.~Y. Wang, and T.~Chen.
\newblock An experimental study of particle damping for beams and plates.
\newblock \emph{J. Vib. Acoust.}, 126:\penalty0 141--148, 2004.
\newblock \doi{10.1115/1.1640354}.

\bibitem[Naeim et~al.(2011)Naeim, Lew, Carpenter, Youssef, Rojas, Saragoni, and
  Adaros]{Naeim}
F.~Naeim, M.~Lew, L.~D. Carpenter, N.~F. Youssef, F.~Rojas, G.~R. Saragoni, and
  M.~S. Adaros.
\newblock Performance of tall buildings in {S}antiago, {C}hile during the 27
  {F}ebruary 2010 offshore {M}aule, {C}hile earthquake.
\newblock \emph{Struct. Design Tall Spec. Build.}, 20:\penalty0 1--16, 2011.
\newblock \doi{10.1002/tal.675}.

\bibitem[Lu et~al.(2012)Lu, Lu, Lu, and Masri]{LU20122007}
Z.~Lu, X.~Lu, W.~Lu, and S.~F. Masri.
\newblock Experimental studies of the effects of buffered particle dampers
  attached to a multi-degree-of-freedom system under dynamic loads.
\newblock \emph{J. Sound Vib.}, 331:\penalty0 2007--2022, 2012.
\newblock \doi{10.1016/j.jsv.2011.12.022}.

\bibitem[Zhou et~al.(2021)Zhou, Li, Shi, Luo, and Deng]{ZHOU2021113073}
Y.~Zhou, D.~Li, F.~Shi, W.~Luo, and X.~Deng.
\newblock Experimental study on mechanical properties of the hybrid lead
  viscoelastic damper.
\newblock \emph{Eng. Struct.}, 246:\penalty0 113073, 2021.
\newblock \doi{10.1016/j.engstruct.2021.113073}.

\bibitem[Salue{\~n}a et~al.(1998)Salue{\~n}a, Esipov, P{\"o}schel, and
  Simonian]{SPIE98}
C.~Salue{\~n}a, S.~E. Esipov, T.~P{\"o}schel, and S.~S. Simonian.
\newblock Dissipative properties of granular ensembles.
\newblock \emph{SPIE: Smart Structures and Materials 1998: Passive Damping and
  Isolation}, 3327:\penalty0 23--31, 1998.
\newblock \doi{10.1117/12.310696}.

\bibitem[Salue{\~n}a et~al.(1999{\natexlab{a}})Salue{\~n}a, Esipov, Rosenkranz,
  and Panossian]{SPIE99}
C.~Salue{\~n}a, S.~E. Esipov, D.~Rosenkranz, and H.~V. Panossian.
\newblock Modeling of arrays of passive granular dampers.
\newblock \emph{SPIE: Smart Structures and Materials: Passive Damping and
  Isolation}, 3672:\penalty0 32, 1999{\natexlab{a}}.
\newblock \doi{10.1117/12.349800}.

\bibitem[Salue{\~n}a et~al.(1999{\natexlab{b}})Salue{\~n}a, P\"oschel, and
  Esipov]{saluena1999}
C.~Salue{\~n}a, T.~P\"oschel, and S.~E. Esipov.
\newblock Dissipative properties of vibrated granular materials.
\newblock \emph{Phys. Rev. E}, 59:\penalty0 4422--4425, 1999{\natexlab{b}}.
\newblock \doi{10.1103/PhysRevE.59.4422}.

\bibitem[Wang et~al.(2015)Wang, Liu, Shan, and He]{wang2015}
X.~Wang, X.~Liu, Y.~Shan, and T.~He.
\newblock Design, simulation and experiment of particle dampers attached to a
  precision instrument in spacecraft.
\newblock \emph{J. Vibroengineering}, 17:\penalty0 1605--1614, 2015.
\newblock \doi{10.1177/16878140211044923}.

\bibitem[Hashemnia(2021)]{Hashemnia2021}
K.~Hashemnia.
\newblock Effect of particle size and media volume fraction on the vibration
  attenuation of a thin-walled beam containing granular media.
\newblock \emph{Soil Dyn. Earthq. Eng.}, 147:\penalty0 106816, 2021.
\newblock \doi{10.1016/j.soildyn.2021.106816}.

\bibitem[Els(2011)]{els2011}
Daniel N.~J. Els.
\newblock Damping of rotating beams with particle dampers: experimental
  analysis.
\newblock \emph{AIAA J.}, 49:\penalty0 2228--2238, 2011.
\newblock \doi{10.2514/1.J050984}.

\bibitem[Papalou and Masri(1998)]{Papalou_1998}
A.~Papalou and S.~F. Masri.
\newblock An experimental investigation of particle dampers under harmonic
  excitation.
\newblock \emph{J. Vib. Control}, 4:\penalty0 361--379, 1998.
\newblock \doi{10.1177/107754639800400402}.

\bibitem[Chen et~al.(2001)Chen, Mao, Huang, and Wang]{Mao2001}
T.~Chen, K.~Mao, X.~Huang, and M.~Y. Wang.
\newblock Dissipation mechanisms of nonobstructive particle damping using
  discrete element method.
\newblock \emph{SPIE: Smart Structures and Materials: Passive Damping and
  Isolation}, 4331:\penalty0 294, 2001.
\newblock \doi{10.1117/12.432713}.

\bibitem[Pourtavakoli et~al.(2016)Pourtavakoli, Parteli, and
  P{\"o}schel]{pourtavakoli2016}
H.~Pourtavakoli, E.~J.~R. Parteli, and T.~P{\"o}schel.
\newblock Granular dampers: does particle shape matter?
\newblock \emph{New J. Phys.}, 18:\penalty0 073049, 2016.
\newblock \doi{10.1088/1367-2630/18/7/073049}.

\bibitem[Kollmer et~al.(2013)Kollmer, Sack, Heckel, and
  P{\"o}schel]{kollmer2013}
J.~E. Kollmer, A.~Sack, M.~Heckel, and T.~P{\"o}schel.
\newblock Relaxation of a spring with an attached granular damper.
\newblock \emph{New J. Phys.}, 15, 2013.
\newblock \doi{10.1088/1367-2630/15/9/093023}.

\bibitem[Meyer and Seifried(2021)]{meyer2021}
N.~Meyer and R.~Seifried.
\newblock Toward a design methodology for particle dampers by analyzing their
  energy dissipation.
\newblock \emph{Comput. Part. Mech.}, 8:\penalty0 681--699, 2021.
\newblock \doi{10.1007/s40571-020-00363-0}.

\bibitem[Ferreyra et~al.(2021{\natexlab{a}})Ferreyra, G{\'o}-Paccapelo, Suarez,
  and Pugnaloni]{ferreyra2021avoiding}
M.~V. Ferreyra, J.~M. G{\'o}-Paccapelo, R.~Suarez, and L.~A. Pugnaloni.
\newblock Avoiding chaos in granular dampers.
\newblock \emph{EPJ Web of Conferences}, 249:\penalty0 15003,
  2021{\natexlab{a}}.
\newblock \doi{10.1051/epjconf/202124915003}.

\bibitem[Wang et~al.(2016)Wang, Liu, Tian, and Tang]{wang2016}
Y.~Wang, B.~Liu, A.~Tian, and W.~Tang.
\newblock Experimental and numerical investigations on the performance of
  particle dampers attached to a primary structure undergoing free vibration in
  the horizontal and vertical directions.
\newblock \emph{J. Sound Vib.}, 371:\penalty0 35--55, 2016.
\newblock \doi{10.1016/j.jsv.2016.01.056}.

\bibitem[Zhang et~al.(2016)Zhang, Chen, Wang, and Fang]{ZHANG2016}
K.~Zhang, T.~Chen, X.~Wang, and J.~Fang.
\newblock Rheology behavior and optimal damping effect of granular particles in
  a non-obstructive particle damper.
\newblock \emph{J. Sound Vib.}, 364:\penalty0 30--43, 2016.
\newblock \doi{10.1016/j.jsv.2015.11.006}.

\bibitem[Ferreyra et~al.(2021{\natexlab{b}})Ferreyra, Baldini, Pugnaloni, and
  Job]{ferreyra2021}
M.~V. Ferreyra, M.~Baldini, L.~A. Pugnaloni, and S.~Job.
\newblock Effect of lateral confinement on the apparent mass of granular
  dampers.
\newblock \emph{Granul. Matter}, 23:\penalty0 45, 2021{\natexlab{b}}.
\newblock \doi{10.1007/s10035-021-01090-w}.

\bibitem[Lu et~al.(2014)Lu, Lu, Jiang, and Masri]{lu2014}
Z.~Lu, X.~Lu, H.~Jiang, and S.~F. Masri.
\newblock Discrete element method simulation and experimental validation of
  particle damper system.
\newblock \emph{Engineering Computations}, 31:\penalty0 810--823., 2014.
\newblock \doi{10.1108/EC-08-2012-0191}.

\bibitem[S{\'a}nchez et~al.(2012)S{\'a}nchez, Rosenthal, and
  Pugnaloni]{SANCHEZ20124389}
M.~S{\'a}nchez, G.~Rosenthal, and L.~A. Pugnaloni.
\newblock Universal response of optimal granular damping devices.
\newblock \emph{J. Sound Vib.}, 331:\penalty0 4389--4394, 2012.
\newblock \doi{10.1016/j.jsv.2012.05.001}.

\bibitem[P\"oschel and Schwager(2005)]{Algo}
T.~P\"oschel and T.~Schwager.
\newblock \emph{Computational Granular Dynamics: Models and Algorithms}.
\newblock Springer, 2005.
\newblock \doi{10.1007/3-540-27720-X}.

\bibitem[Matuttis and Chen(2014)]{Matuttis:2014}
H.-G. Matuttis and J.~Chen.
\newblock \emph{Understanding the Discrete Element Method: Simulation of
  Non-Spherical Particles for Granular and Multi-Body Systems}.
\newblock Wiley, 2014.
\newblock \doi{10.1002/9781118567210}.

\bibitem[Hertz(1881)]{Hertz:1881}
H.~Hertz.
\newblock Ueber die {B}er{\"u}hrung fester elastischer {K}{\"o}rper.
\newblock \emph{J. reine und angewandte Math.}, 92:\penalty0 156--171, 1881.
\newblock \doi{10.1515/crll.1882.92.156}.

\bibitem[Brilliantov et~al.(1996)Brilliantov, Spahn, Hertzsch, and
  P\"oschel]{BSHP}
N.~V. Brilliantov, F.~Spahn, J.-M. Hertzsch, and T.~P\"oschel.
\newblock Model for collisions in granular gases.
\newblock \emph{Phys. Rev. E}, 53:\penalty0 5382--5392, 1996.
\newblock \doi{10.1103/PhysRevE.53.5382}.

\bibitem[M\"uller and P\"oschel(2011)]{MuellerPoeschel:2011}
P.~M\"uller and T.~P\"oschel.
\newblock Collision of viscoelastic spheres: Compact expressions for the
  coefficient of normal restitution.
\newblock \emph{Phys. Rev. E}, 84:\penalty0 021302, 2011.
\newblock \doi{10.1103/PhysRevE.84.021302}.

\bibitem[Bai et~al.(2009)Bai, Keer, Wang, and Snurr]{Bai2009}
X.-M. Bai, L.~M. Keer, Q.~J. Wang, and R.~Q. Snurr.
\newblock Investigation of particle damping mechanism via particle dynamics
  simulations.
\newblock \emph{Granul. Matter}, 11:\penalty0 417, 2009.
\newblock \doi{10.1007/s10035-009-0150-6}.

\bibitem[Bannerman et~al.(2011)Bannerman, Kollmer, Sack, Heckel, Mueller, and
  P\"oschel]{Bannerman2011}
M.~N. Bannerman, J.~E. Kollmer, A.~Sack, M.~Heckel, P.~Mueller, and
  T.~P\"oschel.
\newblock Movers and shakers: Granular damping in microgravity.
\newblock \emph{Phys. Rev. E}, 84:\penalty0 011301, 2011.
\newblock \doi{10.1103/PhysRevE.84.011301}.

\bibitem[{\v{S}}milauer et~al.(2021){\v{S}}milauer, Angelidakis, Catalano,
  Caulk, Chareyre, Ch{\`e}vremont, Dorofeenko, Duriez, Dyck, Eli{\'a}{\v{s}},
  Er, Eulitz, Gladky, Guo, Jakob, Kneib, Kozicki, Marzougui, Maurin, Modenese,
  Pekmezi, Scholt{\`e}s, Sibille, Str{\'a}nsk{\'y}, Sweijen, Thoeni, and
  Yuan]{vsmilauer2021yade-3}
V.~{\v{S}}milauer, V.~Angelidakis, E.~Catalano, R.~Caulk, B.~Chareyre,
  W.~Ch{\`e}vremont, S.~Dorofeenko, J.~Duriez, N.~Dyck, Eli{\'a}{\v{s}}, B.~Er,
  A.~Eulitz, A.~Gladky, N.~Guo, C.~Jakob, F.~Kneib, J.~Kozicki, D.~Marzougui,
  R.~Maurin, C.~Modenese, G.~Pekmezi, L.~Scholt{\`e}s, L.~Sibille,
  J.~Str{\'a}nsk{\'y}, T.~Sweijen, K.~Thoeni, and C.~Yuan.
\newblock \emph{Yade Documentation}.
\newblock 2021.
\newblock \doi{10.5281/zenodo.5705394}.

\bibitem[Opsomer et~al.(2011)Opsomer, Ludewig, and Vandewalle]{Opsomer:2011}
E.~Opsomer, F.~Ludewig, and N.~Vandewalle.
\newblock Phase transitions in vibrated granular systems in microgravity.
\newblock \emph{Phys. Rev. E}, 84:\penalty0 051306, 2011.
\newblock \doi{10.1103/PhysRevE.84.051306}.

\end{thebibliography}

\end{document}